\documentclass[aps,pre,preprint,amsmath,amssymb]{revtex4}

\usepackage{graphicx}
\usepackage{soul}
                                                                                                                             
\begin{document}
                                                                                                                             
\bibliographystyle{apsrev}
                                                                                                                             
\title{Water-mediated interactions between hydrophobic and ionic species in cylindrical nanopores}

\author{S. Vaitheeswaran$^{1}$, G. Reddy$^{1}$ and D. Thirumalai$^{1,2,*}$,\\
$^1$Biophysics Program, Institute for Physical Science and Technology,\\
$^2$Department of Chemistry and Biochemistry,\\
University of Maryland, College Park, MD 20742\\
Email: {\tt thirum@umd.edu}}
                                                                                                                             
\date{\today}
                                                                                                                             
                                                                                                                             
\begin{abstract}
We use Metropolis Monte Carlo and umbrella sampling to calculate the free energies of interaction
of two methane molecules and their charged derivatives in cylindrical water-filled pores.
Confinement strongly alters the interactions between the nonpolar solutes, and completely eliminates
the solvent separated minimum (SSM) that is seen in bulk water.
The free energy profiles show that the methane molecules are either in
contact or at separations corresponding to the diameter and the length of the cylindrical pore.
Analytic calculations that estimate the entropy of the solutes, which are solvated at the pore surface,
qualitatively explain the shape of the free energy profiles.
Adding charges of opposite sign and magnitude $0.4e$ or $e$ (where $e$ is the electronic
charge) to the methane molecules decreases their tendency for surface solvation and restores the SSM.
We show that confinement induced ion-pair formation occurs whenever $l_B/D \sim O(1)$,
where $l_B$ is the Bjerrum length, and $D$ is the pore diameter.
The extent of stabilization of the SSM increases with ion charge density as long as $l_B/D < 1$.
In pores with $D \le 1.2$ nm, in which the water is strongly layered,
increasing the charge magnitude from $0.4e$ to $e$ reduces the stability of the SSM.
As a result, ion-pair formation which occurs with negligible probability in the bulk, is promoted.
In larger diameter pores that can accomodate a complete hydration layer around the solutes,
the stability of the SSM is enhanced.
\end{abstract}

\maketitle

{\bf Introduction:}

\bigskip

Confinement effects on the properties of water and water-mediated interactions between solutes
are receiving increasing attention because of their relevance to a variety of problems in
physical chemistry \cite{Gelb_Gubbins_RepProgPhys99,Truskett_JCP01} and biophysics
\cite{Thirumalai_Lorimer_ARBBS01}.
In this study, we investigate the interactions between methane molecules and their charged
derivatives in water-filled cylindrical nanopores.
The interactions between these moieties are assessed in terms of the potential of mean force (PMF), 
which is the reversible work required to bring the solutes from infinity to $r$, the
distance between the solutes.
We are interested in such a geometry because, quantitative calculations of well defined
systems might provide insights into more complex situations, such as the conformations
of a polypeptide chain inside the cylindrical tunnel of the ribosome \cite{O'Brien_Thirumalai_NanoLett08}
and peptide transport through biological channels such as $\alpha$-hemolysin
\cite{Movileanu_Bayley_BiophysJ05}.
Our results are also applicable to other situations where the conformations of biopolymers are important, 
such as transport of nucleic acids through carbon nanotubes \cite{Yeh_Hummer_PNAS04} and synthetic nanopores
\cite{Heng_Timp_BiophysJ06}, and in nanopore sequencing of nucleic acids \cite{Schmidt_JMatChem05,
Deamer_Branton_ACR02}. Other situations of interest may include studies of adsorbed solutes
in porous materials like zeolites.

In a previous study, we showed that hydrophobic interactions in spherical, nanometer-sized
water droplets \cite{Vaitheeswaran_JACS06} differ significantly from those in bulk water.
The PMF is independent of the size of the confining droplet and, in contrast to the bulk case,
shows only a single minimum at contact; the solvent separated minimum (SSM) is completely absent.
Interactions between oppositely charged solutes in spherical droplets show that the extent of
solvation is determined by the charge density of the ion \cite{Vaitheeswaran_JACS06}.
Here, we investigate the effect of confinement in cylindrical nanopores on the interactions
between methane molecules and their charged derivatives \cite{Mountain_Thirumalai_JACS03}. 
We show that, in a nanopore filled with water at the same density ($\rho$) as in the bulk,
the SSM in the interaction free energy between methanes is absent.
The contact minimum is metastable with respect to two other minima, at separations corresponding
to the diameter or the length of the pore. 
To understand the effect of confinement in a cylindrical volume on charged solutes, we also
calculate the free energies of interaction between the ions M$_{q+}$ and M$_{q-}$.
One of the methanes has a positive charge $q^+$ while the other carries a negative charge. 
The results illustrate how the balance between solute-solute and solute-solvent interactions depends
on the charge density and the degree of confinement.
We conclude by discussing the implications of our results for protein stability in confinement.

\bigskip

\textbf{Methods:}

\bigskip

\textbf{Models.}
We investigate the interactions between two methane molecules in cylindrical, water-filled
nanopores using Metropolis Monte Carlo \cite{Metropolis_JCP53} simulations. 
We use the TIP3P model for water \cite{Jorgensen_Klein_JCP83}, and a unified atom representation for methane.
The Lennard-Jones (LJ) interaction parameters involving the solutes \cite{Kalra_Hummer_Garde_JPCb04}
are given in Table~\ref{tab:LJ_parms}.
Model ions M$_{q+}$ and M$_{q-}$ are created by adding charges of magnitude $q^+ = |q^-| = 0.4e$
or $e$ to the methane spheres, where $e$ is the electronic charge \cite{Wallqvist_Thirumalai_JACS98}. 

\textbf{Simulation Details.}
Simulations are performed at constant number of molecules $N$,
total volume $V$ and temperature $T=298$ K, and thus sample the canonical ensemble.
We assume that the effective volume available to the water molecules is
\begin{equation}
  V_{\mathrm{eff}} = V - N_sV_s
\end{equation}
where $N_s=2$ is the number of solute molecules in the pore, and $V_s = (4/3)\pi \sigma_{MO}^3$,
is the excluded volume due to each solute. 
For a chosen number of water molecules, $N_w$, and pore diameter $D$, the pore
length $L$ is calculated using
\begin{equation}
  V_{\mathrm{eff}} = \pi \left(\frac{D}{2}\right)^2 L
\end{equation}
and
\begin{equation}
  \rho _w = \frac{N_w \ m_w}{V_{\mathrm{eff}}}
\end{equation}
where $\rho_w=997$~kg/m$^3$ is the water density at 298 K and 1 atm pressure, and $m_w$ is the
molecular weight of water.
The pore dimensions and the corresponding $N_w$ are listed in Table~\ref{tab:N_w}.
The potential energy due to the pore walls in the cylindrical polar coordinates $\xi$ and $z$ is given by
\begin{eqnarray}
U_1(\xi) &=& \left\{ \begin{array}{ll} 0 & \xi \leq D/2 \\
   10^{12} \hspace{0.5cm} \mathrm{kJ/mol} \hspace{1.7cm} & \xi > D/2 \end{array} \right. \hspace{2.8cm} \rm{and}\\
U_2(z) &=& \left\{ \begin{array}{ll} 0 & |z| \leq L/2 \\
               10^{12} \hspace{0.5cm} \mathrm{kJ/mol} \hspace{1.7cm} & |z| > L/2 \end{array} \right.
\end{eqnarray}
In order to simulate confinement effects we do not use periodic boundaries.
Electrostatic and Lennard-Jones interactions are evaluated without a cutoff. 

Monte Carlo trial moves consisted of a random translation for both solute and solvent molecules
and a random rotation for solvent molecules only. 
In addition, we also used collective translations and collective flips, $z \rightarrow -z$
(the z-axis being parallel to the pore axis), of the methanes.
The moves were accepted according to the Metropolis criterion \cite{Metropolis_JCP53}.
We used umbrella sampling to calculate the free energy of interaction between 
two methane molecules, $-k_BT \log P(r)$, where $k_B$ is Boltzmann's constant and $P(r)$ is the
probability of finding the two solutes a distance $r$ apart.
Because of the quasi one-dimensional nature of the confinement when $r \gg D$, we do not subtract
the free energy contribution $-2k_BT \log r$ which arises from the increase in phase space
proportional to $r^2$ in spherically symmetric systems.
Therefore, these profiles cannot be compared directly to PMFs in bulk.
The biasing potentials are harmonic, $U_{B}=0.5k(r-r_0)^2$, where $r_0$ is the
center of each window and the spring constant $k$ is chosen to be 5 kcal/(mol-\r{A}$^2$).
Window centers are 0.1 nm apart in all calculations. The data is unbiased
and the free energies are calculated using the weighted histogram analysis method (WHAM)
\cite{Kumar_JCompChem92} with code from Crouzy et~al.~\cite{Crouzy_JCompChem99}.
All PMFs and free energies in this work are only determined to within
an additive constant; the free energy at contact is arbitrarily chosen to be zero.

\bigskip

\textbf{Results and Discussion:}

\bigskip

{\bf Equilibrium structure of confined water.} Fig.~\ref{fig:rad_dens_prof}a shows
the local solvent density divided by the average density $\rho_0 = N_w/V_{\mathrm{eff}}$
for a pore with $D=0.8$ nm, and number of water molecules $N_w$ varying from 10 to 128.
For $N_w=10-32$, the reduced density is less than 1 for all values of $\xi$, implying that the excluded
volume due to the solutes has been overestimated. For $N_w=10$ the solvent has no structure.
As $N_w$ and $L$ increase, two concentric water layers become progressively well defined.

Fig.~\ref{fig:rad_dens_prof}b shows the radial density profiles in nanopores with varying $D$ and $L$
and $N_w=128$. For the narrowest pores with $D=0.8$ and 1.2 nm, two sharply defined concentric
layers are seen. The volume available to the waters in the $D=0.8$ pore is close to the volume accessible
to water molecules in the periodically replicated channel with repulsive walls and diameter 0.9 nm
simulated by Lynden-Bell and Rasaiah \cite{Lynden-Bell_Rasaiah_JCP96}; correspondingly, solvent structure
is similar in both cases. With increasing $D$, the layering is lost and the solvent becomes more bulk-like in
the interior with substantial depletion at the surface ($\xi \approx D/2$).

In Fig.~\ref{fig:Mu}a we plot the probability distributions, $P(\mu_z/\mu)$ for pores with
$D=0.8$ nm and varying $N_w$.
The dipole moment of a single water molecule is $\mu$, and $\mu_z$ is its value along the z-axis.
Fig.~\ref{fig:Mu}b shows $P(\mu_z/\mu)$ for $N_w=128$ and varying values of $D$.
In Fig.~\ref{fig:Mu}a, individual water dipoles are seen to be preferentially
aligned with the pore axis, pointing either ``up'' or ``down''.
The extent of  ordering increases with increasing pore length.
Fig.~\ref{fig:Mu}b shows that, as the diameter increases from 0.8 nm to 2.0 nm, the strong
ordering in water dipoles decreases and is altogether absent for the widest pore with $D=2$ nm.
However, there is no net dipole moment in the pores.
The dashed lines show the probability distributions for the total dipole moment of all water molecules in a
nanopore with $D=0.8$ nm and $N_w=128$ (Fig.~\ref{fig:Mu}a) and $D=2.0$ nm and $N_w=128$ (Fig.~\ref{fig:Mu}b).
The distribution in other nanopores is similar (data not shown).
Thus, Figs.~\ref{fig:rad_dens_prof} and \ref{fig:Mu} show that solvent structure deviates the
most from bulk behavior in long and narrow pores.

{\bf Free energies of interaction between methane molecules.} 
Contours of methane positional probabilities in nanopores, with $N_w=128$ and $D=0.8$ nm in
Fig.~\ref{fig:contour}a and for $D=2.0$ nm in Fig.~\ref{fig:contour}b show that the solutes
are most likely to be found at the periphery of the pore where water hydrogen bonds are broken
and water density is significantly lower (Fig.~\ref{fig:rad_dens_prof}b).
The effective interaction between these nonpolar solutes reflects the nature of the confining volume.
Fig.~\ref{fig:PMF_d0.8} shows free energies of interaction of two methane molecules in pores with
$D=0.8$ nm and $L \in [1.27,2.59]$ nm.
In all these pores, the solutes are most stable in contact with each other at $r \approx 0.4$ nm.
The SSM that is seen in the methane-methane PMF in bulk water \cite{Pangali_Berne_JCP79,Shimizu_Chan_JCP00},
is conspicuously absent.
In pores with $D=0.8$ nm, a distant minimum occurs at $r \approx L$.
In all pores with $L \in [1.27,2.59]$ nm (Fig.~\ref{fig:PMF_d0.8}), the distant minimum is less favorable
than the contact minimum by about 3 kJ/mol ($\sim1.2 k_BT$ at $T=298$ K).
The height of the barrier between the two states increases from $\sim3.2$ kJ/mol to $\sim8$ kJ/mol with
increasing pore length in this range, due to the greater
free energy cost of disrupting the water column in the longer pores that are more ordered
(Figs.~\ref{fig:rad_dens_prof} and \ref{fig:Mu}).
In Fig.~\ref{fig:Trans_entropy} we plot free energy profiles (solid lines) in
nanopores with $D \in [1.2,2.0]$ nm and $N_w=128$. 
In all these pores, the contact minimum occurs at $r=0.4$ nm.
A distant minimum at $r \approx$ max\{$D,L$\} (see Table~\ref{tab:N_w}), is marginally more stable
than the contact minimum by about $0.5-0.7$ kJ/mol.
There is another distant minimum at $r \approx$ min\{$D,L$\}.
In the case of the narrowest pores ($D=0.8$ nm), the minimum
corresponding to $D$ almost merges with the one at contact and appears as a shoulder at 0.7 nm. 
With increasing $D$, the barrier between the contact minimum and the distant minimum at
$r \approx$ min\{$D,L$\} decreases, because the greater
water depletion at the pore surface (Fig.~\ref{fig:rad_dens_prof}b) lowers the free energy barriers
and makes the states with intermediate $r$ less unfavorable.

\textbf{Solute translational entropy due to surface solvation is a major contribution to the free energy.}
The free energy profiles in Fig.~\ref{fig:PMF_d0.8} and \ref{fig:Trans_entropy} can be qualitatively
understood on the basis of the surface solvation of nonpolar solutes in confinement.
At the hydrophobic pore boundary, the structure of water is disrupted due to the breakage of hydrogen bonds.
This gives rise to a depletion zone of low solvent density at the boundary.
At equilibrium, hydrophobic species such as methane will be localized at these extended cavities at the water surface.
This result is further supported by quantum mechanical simulations which find that even an alanine zwitterion
in an isolated water droplet \cite{Degtyarenko_JPCb07} is preferentially solvated at the surface,
with the nonpolar side chain exposed to air.
The entropic preference of nonpolar solutes for the surface of water-filled, periodically
replicated, cylindrical channels has also been reported in other computational studies
\cite{Lynden-Bell_Rasaiah_JCP96}.
The SSM seen in the PMF between two methane-sized hydrophobic solutes in bulk
water can be attributed to the presence of a single water molecule between the solutes, which is
hydrogen bonded to other waters.
In the solvent depleted zone near the walls of the pore where water hydrogen bonds are broken,
this conformation is strongly unfavorable.
Therefore, in hydrophobic confinement the SSM is absent.

The contribution of the solute translational entropy to the interaction free energies can be calculated approximately, 
assuming that the hydrophobic species are hard spheres that are strictly confined to the surface of the nanopores.
The translational entropy at a solute separation $r$, is proportional to the length of a contour defined
by the intersection of a sphere of radius $r$ and a capped cylinder with the dimensions of the pore,
with the sphere center located on the cylinder boundary (see Appendix and Fig.~\ref{fig:cyl_sph_inter}).
The contribution of this purely geometric term to the methane-methane free energy, $-T \Delta S^{\mathrm{A}}$,
is plotted in black for three different pores in Fig.~\ref{fig:Trans_entropy}.
As with the total free energy, $-T \Delta S^{\mathrm{A}}$ is only determined to within
an additive constant, and is vertically aligned with the first distant minimum in the free energy at
$r \approx$ min\{$D,L$\}.
Fig.~\ref{fig:Trans_entropy} shows that $-T \Delta S^{\mathrm{A}}$ qualitatively captures the broad
curvature of the free energy profiles, especially for the wider pores.
Clearly, the solute translational entropy favors the first distant minimum in the free energy, while the
solvent drives the solutes to either minimize or maximize their separation.
We note that the solutes are not localized exactly at the pore surface
(Fig.~\ref{fig:contour}), due to attractive van der Waals interactions with the solvent.
The first distant minimum therefore occurs at $r <$ min\{$D,L$\}.

{\bf Interaction between charged solutes: Extreme confinement induces ion-pair formation.} 
In spherical water droplets \cite{Vaitheeswaran_JACS06}, the tendency of solutes for surface or interior solvation
depends on their charge density ($\zeta$) and the curvature of the confining surfaces.
For a given solute size and charge magnitude, anions have a greater preference for the surface than cations.
The tendency for surface solvation in periodically replicated cylindrical channels also correlates
with $\zeta$.
In simulations of monovalent cations and anions in a channel of diameter 0.3 nm
\cite{Lynden-Bell_Rasaiah_JCP96}, smaller ions were found close to the axis of the channel, while larger
ones with the same charge were more likely to be located closer to the wall.

The reduced preference of charged solutes for surface solvation, as compared to nonpolar solutes,
is reflected in the free energies of interaction between model ions M$_{q+}$ and
M$_{q-}$ of charge magnitude $q^+ = |q^-| = 0.4e$ (Fig.~\ref{fig:PMF_q0.4}).
In all the pores studied here, there is a contact minimum at 0.37 nm and a SSM at
$\sim0.67$ nm, with a difference of $2-4$ kJ/mol between the two states. Relative to the contact state,
the solvent separated state becomes more stable with increasing pore length at $D=0.8$ nm
(Fig.~\ref{fig:PMF_q0.4}a), and also with increasing $D$ at fixed $N_w=128$
(Fig.~\ref{fig:PMF_q0.4}b).

Fig.~\ref{fig:PMF_q1.0} shows the same free energies evaluated for solute charge magnitudes of $e$.
Comparing Figs.~\ref{fig:PMF_q0.4}a and ~\ref{fig:PMF_q1.0}a, we see that increasing the solute charge
in pores of $D=0.8$ nm destabilizes the SSM.
Similar findings hold good for the nanopore with $D=1.2$ nm (Figs.~\ref{fig:PMF_q0.4}b and ~\ref{fig:PMF_q1.0}b).
However, for the wider pores with $D=1.6$ and 2.0 nm, the opposite trend is observed; increasing solute charge
makes the SSM more stable relative to the contact minimum. 

In the solvent separated state, the two solutes are stabilized in individual hydration shells with a
single water molecule between them that is hydrogen bonded to other water molecules. From M$_{q\pm}$-water
radial distribution functions (data not shown), we infer that the first hydration shell around the solvent
separated solute pair with charge magnitude 0.4$e$, has a diameter of $\approx 1.6$ nm. For the solutes
with a charge magnitude of $e$, the corresponding hydration shell has a diameter of $\approx 1.3$ nm, with
a second hydration shell of diameter $\approx 1.7$ nm. Thus, pores with $D \le 1.2$ nm are too narrow to
accomodate even a single hydration shell around the solvent separated solute pair with a charge magnitude
of $e$.
The direct solute-solute coulombic attraction is therefore stronger, and thus in these pores, ion-pair
formation is promoted, especially when the charge increases from 0.4$e$, to $e$.
The opposite happens in pores with $D \ge 1.6$ nm in which the solute-solvent interaction energy is
more favorable.
Clearly, as $D \rightarrow \infty$ one recovers the bulk behavior in which the ions are fully hydrated.

\bigskip

\textbf{Conclusions:}

\bigskip

The mutual interaction between solute molecules
confined in a cylindrical pore is strongly influenced by the presence of the confining boundary.
Solute-solute interactions differ drastically from those in bulk solvent.
Hydrophobic interactions are altered more than electrostatic interactions.
Nonpolar solutes, such as the methane-methane pair, are driven to the pore surface to maximize solvent entropy.
Consequently, the solvent separated state that is seen in bulk water is absent in nanopores.
Methane-methane free energy profiles show one contact minimum and two distant minima, at separations
corresponding to the diameter and the length of the pore.
The broad curvature of the free energy profiles is shown to be largely determined by the translational
entropy of the solutes at the pore surface.
The solvent drives the hydrophobic solutes either to contact or to the largest possible separation in the pore.
The intervening minimum arises due to the greater solute entropy at $r \approx$ min\{$D,L$\}.
The stability of the distant minima, relative to the contact minimum, depends on the pore dimensions.

Model ions M$_{q+}$ and M$_{q-}$, created by adding charges of magnitude $q^+ = |q^-| = 0.4e$ or $e$ to
the methanes, have a more favorable energy of interaction with the solvent.
These charged solutes are solvated in the pore interior, away from the surface, and their interaction free
energy shows a SSM.
For a given charge density, this state becomes more stable relative to the contact state, as the
volume of confinement increases.
Increasing the charge from $0.4e$ to $e$ increases the stability of the solvent separated state for the
pores with $D \ge 1.2$ nm, which can accomodate a complete hydration shell around the ion pair.
In narrower pores, increasing the ion charge destabilizes the SSM.

The interactions between charged species in cylindrical confinement can be rationalized using an interplay
of two length scales.
In the presence of charges, an additional length scale $l_B = z^2 e^2 / \epsilon k_B T$ ($z$ is the valence
and $\epsilon$ is the dielectric constant) plays an important role.
The balance between $l_B$ and $D$ determines the stability of the SSM.
To a first approximation, the stability of the second minimum (Figs.~\ref{fig:PMF_q0.4} and ~\ref{fig:PMF_q1.0})
is determined by $l_B/D$.
As $l_B/D$ decreases at a fixed charge magnitude, the second minimum becomes more pronounced.

The preference of aliphatic groups to be localized at the pore boundary and charged species to be solvated
in the pore interior can be used to infer the stability changes of peptides confined to cylindrical pores.
For a generic amphiphilic sequence it is likely that the reduction in the conformational entropy of the
denatured states, with respect to the bulk, and the propensity of polar and charged residues to be fully
solvated, should compensate for the tendency of hydrophobic species to be pinned at the interface.
Thus, we predict that, confinement in water-filled cylindrical nanopores should enhance the
stability of confined amphiphilic peptides.
Preliminary molecular dynamics \cite{Vaitheeswaran_PNAS08} and Langevin dynamics \cite{O'Brien_Thirumalai_NanoLett08}
simulations support these arguments.
The interactions between charged species M$_{q+}$ and M$_{q-}$ with $q^+ = e$ suggest that ion pair formation,
which is unlikely in the bulk, can be promoted in narrow cylindrical pores.
The simulations (Figs.~\ref{fig:PMF_q0.4} and ~\ref{fig:PMF_q1.0}) show that the interaction between charged
solutes is governed by competing solute-solute and solute-solvent interactions, as well as the charge-density
of the ions \cite{Vaitheeswaran_JACS06}.
The importance of charge density in controlling the stability of folded RNA has been demonstrated previously
\cite{Koculi_Thirumalai_Woodson_JACS07}.

\bigskip

\textbf{Appendix: Estimate of entropy of pinning near the pore surface}

\bigskip

In this Appendix, we evaluate the free energy cost of localizing a spherical hydrophobic species near
the surface of the pore.
We idealize the hydrophobic solutes as point hard spheres that are strictly confined to the surface
of the cylindrical pore.
The free energy of interaction of these particles will be purely entropic, and can be calculated from
the pore geometry.
Consider an ideal particle located at the periphery of the pore, either along the curved length of the
cylinder, or one of the flat end caps.
The probability of finding the other particle at a distance $r$ is proportional to the length of the
contour defined by, the intersection of a cylindrical shell with the dimensions of the pore, and a spherical
shell of radius $r$ which is centered on the first particle (see Fig.~\ref{fig:cyl_sph_inter}).
Here, we derive the expression for this contour length.

\bigskip
The equation for the body of a cylinder with its center at the origin and aligned along the $z$-axis is
\begin{equation}\label{cyl}
x^2 + y^2 = a^2 = (D/2)^2,   \  \  -L/2 \leq z \leq L/2 \ .
\end{equation}
Here $a$, $D$ and $L$ are the radius, diameter and length of the cylinder respectively.
The cylinder has caps at $z=L/2$ and $z=-L/2$.
Eq.~(\ref{cyl}) can be parameterized as $x=a \sin\theta$ and $y = a \cos\theta$.
The problem is evaluated in three parts.
   
\section{Sphere center on the body of the cylinder, intersecting the body of the cylinder}
 The equation of a sphere with its center at $(a, 0, -c)$ on the body of the cylinder is
\begin{equation}\label{sph}
(x-a)^2 + y^2 +(z+c)^2= r^2,   \  \  -L/2 \leq c \leq L/2 \ .
\end{equation}
Using the parameterized cylinder equations and solving for $z$ in Eq.~(\ref{sph}) yields
$z=-c \pm \sqrt{r^2 - 2a^2 +2a^2\sin(\theta)}$.
The length of the contour of intersection of the cylinder and the spheres centered on the rim of the
cylinder at $z = -c$ is
 \begin{equation}\label{cont}
 \l_1 = 4 \pi a\int_{0}^{2\pi}\sqrt{(dx/d\theta)^2+(dy/d\theta)^2+(dz/d\theta)^2} \  d\theta, \ -L/2+c \leq  \sqrt{r^2 - 2a^2 +2a^2\sin(\theta)} \leq c+L/2
 \end{equation}
The center of the sphere can be varied along the $z$-axis from -L/2 to L/2.
Dividing by 2 to avoid double counting, the length of this part of the total contour is
\begin{multline}\label{cont2} 
  L_1 (r) = 2 \pi a^2\int_{-L/2}^{L/2} \\
   \left \{ \int_{0}^{2\pi}\sqrt{1+\frac{a^2 \sin^2(\theta)}{r^2-2a^2+2a^2\sin(\theta)}} \  d\theta, -L/2+c \leq  \sqrt{r^2 - 2a^2 +2a^2\sin(\theta)} \leq c+L/2 \right \} \ dc
\end{multline}   

\section{Sphere center on the cylinder cap, intersecting the body of the cylinder}
The equation of a sphere with its center on the cap of the cylinder at $z=-L/2$ and in the $xz$ plane is
\begin{equation}\label{sph2} 
(x-r')^2 + y^2 +(z+\frac{L}{2})^2= r^2,   \  \  0 \leq r' \leq a \ .
\end{equation}
Solving for  $z$ using the parameterized equations of the cylinder yields
$z=-\frac{L}{2} \pm \sqrt{r^2 - a^2 - r'^2 + 2ar'\sin(\theta)}$.
The length of the contour is given by
\begin{equation}\label{cont6}
 \l_1 = 4\int_{0}^{2\pi}\sqrt{(dx/d\theta)^2+(dy/d\theta)^2+(dz/d\theta)^2} \  d\theta, \ 0 \leq  \sqrt{r^2 - a^2 -r'^2 +2ar'\sin(\theta)} \leq L
\end{equation}

The center of the sphere can be any where on the circle on the cap of the cylinder described by
$z = -L/2$ and $x^2 + y^2 = r'^2$.
This gives a multiplicative factor of $2 \pi r'$.
The center of the sphere on the cap can also be varied along the $x$-axis from 0 to $a$.
Therefore, the length of this part of the total contour is
\begin{multline}\label{cont8}
  L_2 (r) = 8 \pi a \int_{0}^{a} r' \\
   \left \{ \int_{0}^{2\pi}\sqrt{1+\frac{r'^2 \sin^2(\theta)}{r^2-a^2-r'^2+2ar'\sin(\theta)}} \  d\theta, 0 \leq  \sqrt{r^2 - a^2 -r'^2+2ar'\sin(\theta)} \leq L \right \} \ dr'
\end{multline}

\section{Sphere center on one of the caps of the cylinder and intersects the second cap}
The equation of a sphere with its center on the $z=-L/2$ cap of the cylinder and in the $xz$ plane is
given by Eq.~(\ref{sph2}).
Since the sphere should intersect the other cap of the cylinder present at $z=L/2$, substituting $z=L/2$ in
Eq.~(\ref{sph2}), we get $(x-r')^2 + y^2 = r^2-L^2,   \  \  0 \leq r' \leq a$.
The equation for the cap of the cylinder at $z=L/2$ is
\begin{equation}
x^2 + y^2 = a^2, \ \  z=L/2  \ .
\end{equation}
The center of the sphere can be any where on the circle on the cap of the cylinder described by
$z = -L/2$ and $x^2 + y^2 = r'^2$.
The center of the sphere on the cap can also be varied along the $x$-axis from 0 to $a$.
Therefore, the length of this part of the total contour is given by
\begin{equation}\label{cont33}
 L_3 (r) = \int_{0}^{a} 4 \pi \theta r' \sqrt{r^2-L^2} \ dr', \ \ \theta = \left\{ \begin{array}{ll} \pi  & \mbox{$\sqrt{r^2-L^2}+r' \leq a,$} \\
\cos^{-1}\left(\frac{r'^2+r^2-L^2-a^2}{2 r' (r^2-L^2)}\right) & \mbox{$\sqrt{r^2-L^2}+r' > a$}
\end{array} \right.
\end{equation}

\bigskip

Finally, the total length of the contour of interest is
\begin{equation}
l_c(r) = L_1(r) + L_2(r) + L_3(r) ,
\end{equation}
where $L_1, L_2$ and $L_3$ are given by Eqs.~(\ref{cont2}), (\ref{cont8}) and (\ref{cont33}) respectively,
and are solved using Mathematica \cite{Mathematica6.0}.
The probability of finding the two particles at a separation $r$ is $P(r) \propto l_c(r)$.
The entropic contribution to the interaction free energy of the two hydrophobic solutes in this study,
can then be approximated (to within an additive constant) as
\begin{equation}
 -T \Delta S^A (r) \approx -k_B T \log l_c(r) \ ,
\end{equation}
where $k_B$ is Boltzmann's constant and $T$ is the temperature.

\section{Acknowledgements}
We thank Ed O'Brien and Riina Tehver for pertinent remarks.
This work was supported in part by a grant from the National Science Foundation (CHE 05-14056)
and the National Institutes of Health (R01 GM076688-05).


\newpage
                                                                                                                             
\begin{table}[!ht]
 \caption{Lennard-Jones \protect\footnotemark[1] parameters between the various sites.}
 \label{tab:LJ_parms}
\setlength{\tabcolsep}{5mm}
 \begin{tabular}{|c|c|c|}
   \hline\hline
   & $\epsilon_{\alpha\beta}$ [kJ/mol] & $\sigma_{\alpha\beta}$ [nm] \\ \hline
methane-methane & 1.23 ($\epsilon_{MM}$) & 0.373 ($\sigma_{MM}$)\\
oxygen-oxygen   & 0.64 ($\epsilon_{OO}$) & 0.315 ($\sigma_{OO}$)\\
methane-oxygen\footnotemark[2]  & 0.887 ($\epsilon_{MO}$) & 0.344 ($\sigma_{MO}$)\\
   \hline\hline
 \end{tabular}
\footnotetext[1]{The Lennard-Jones potential has the generic form $V_{LJ}^{\alpha\beta}(r)
  = 4\epsilon_{\alpha\beta}\left[ \left( \frac{\sigma_{\alpha\beta}}{r} \right)^{12} -
    \left( \frac{\sigma_{\alpha\beta}}{r} \right)^{6} \right]$
  where $\alpha$ and $\beta$ identify the two species, $r$ is the separation distance and $\sigma_{\alpha\beta}$ is
  the collision diameter.}
\footnotetext[2]{The methane-oxygen parameters are obtained from the methane-methane
  \cite{Kalra_Hummer_Garde_JPCb04} and oxygen-oxygen \cite{Jorgensen_Klein_JCP83} parameters
  by applying the Lorentz-Berthelot mixing rules.}
\end{table}

\begin{table}[!ht]
 \caption{Nanopore dimensions for a
  given number of water molecules $N_w$ and number of solute
  molecules $N_m=2$.}
 \label{tab:N_w}
\setlength{\tabcolsep}{10mm}
 \begin{tabular}{|c|c|c|}
   \hline\hline
$N_w$\footnote{Number of water molecules} & $D$ [nm] \footnote{Diameter of the cylinder} &
$L$ [nm] \footnote{Length of the cylinder}\\ \hline 
 10                          & 0.8      & 1.27 \\ \hline
 17                          & 0.8      & 1.69 \\ \hline
 26                          & 0.8      & 2.23 \\ \hline
 32                          & 0.8      & 2.59 \\ \hline
                             & 0.8      & 8.31 \\ \cline{2-3}
                             & 1.2      & 3.69 \\ \cline{2-3}
\raisebox{2.5ex}[0pt]{128}   & 1.6      & 2.08 \\ \cline{2-3}
                             & 2.0      & 1.33 \\ \hline\hline
 \end{tabular}
\end{table}

\newpage
                                                                                                                             
\textbf{Figure Captions:}

\bigskip

\begin{figure}[htbp]
   \centerline{\includegraphics[width=5in]{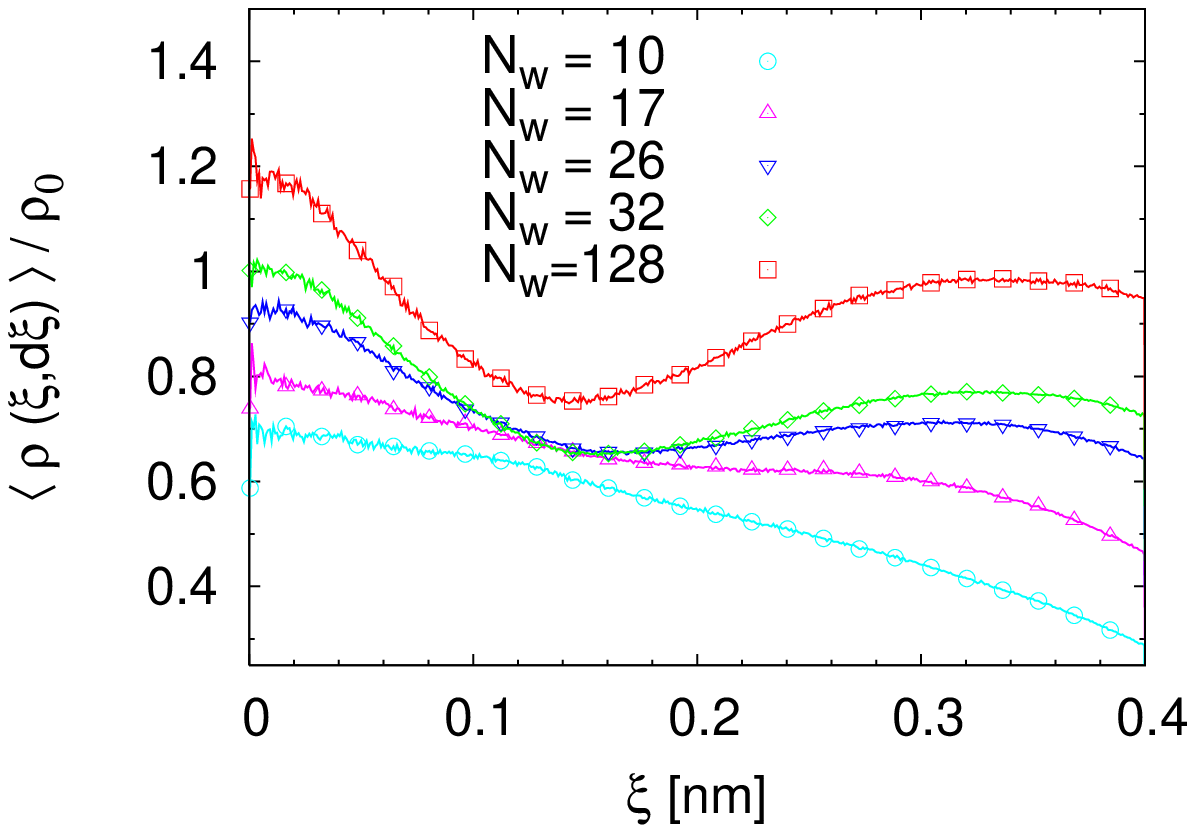}}
   \centerline{\includegraphics[width=5in]{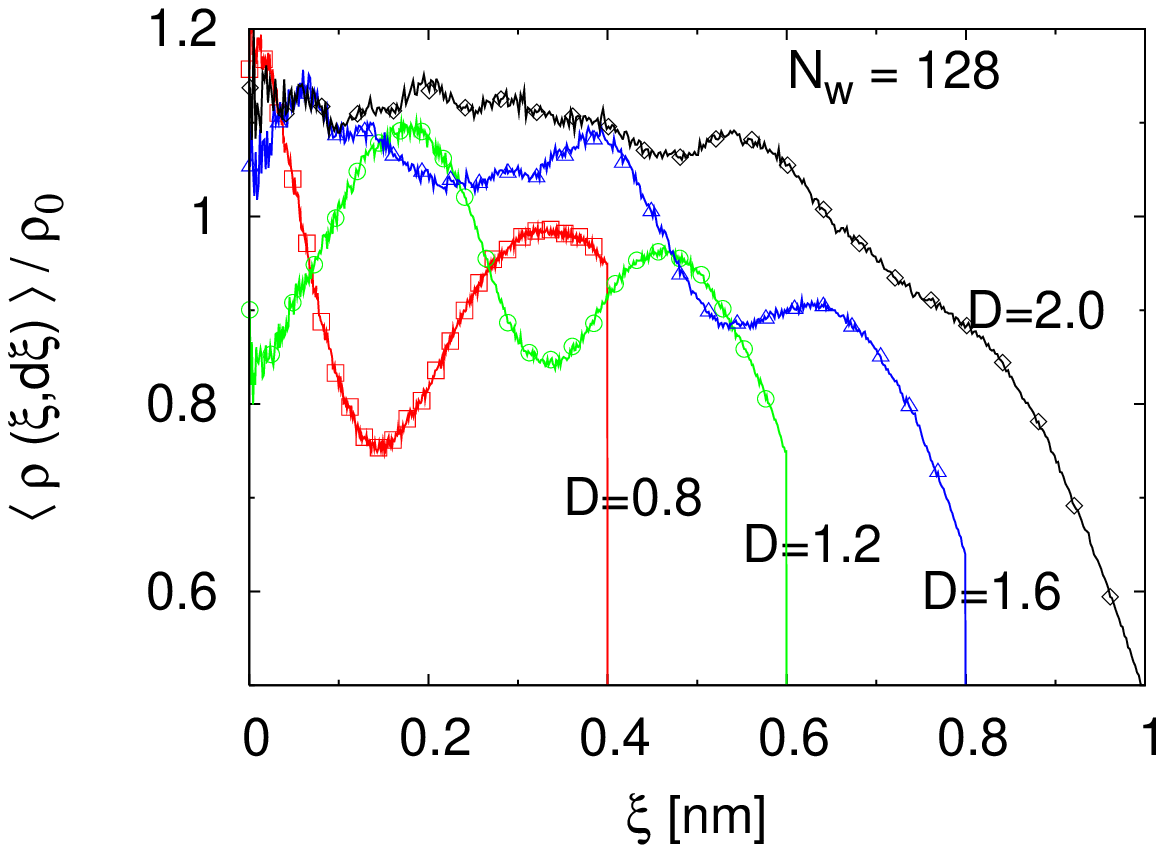}}
   \caption{}
   \label{fig:rad_dens_prof}
 \end{figure}

Fig.~\ref{fig:rad_dens_prof}: (color online) Radial density profiles for water in nanopores
  (a) with diameter $D=0.8$ nm and number of water molecules $N_w$ varying from
  10 to 128 (b) with $N_w=128$ and $D$ varying from 0.8 to 2.0 nm. The length of
  each nanopore is given in Table~\ref{tab:N_w}.

\begin{figure}[htbp]
   \centerline{\includegraphics[width=5in]{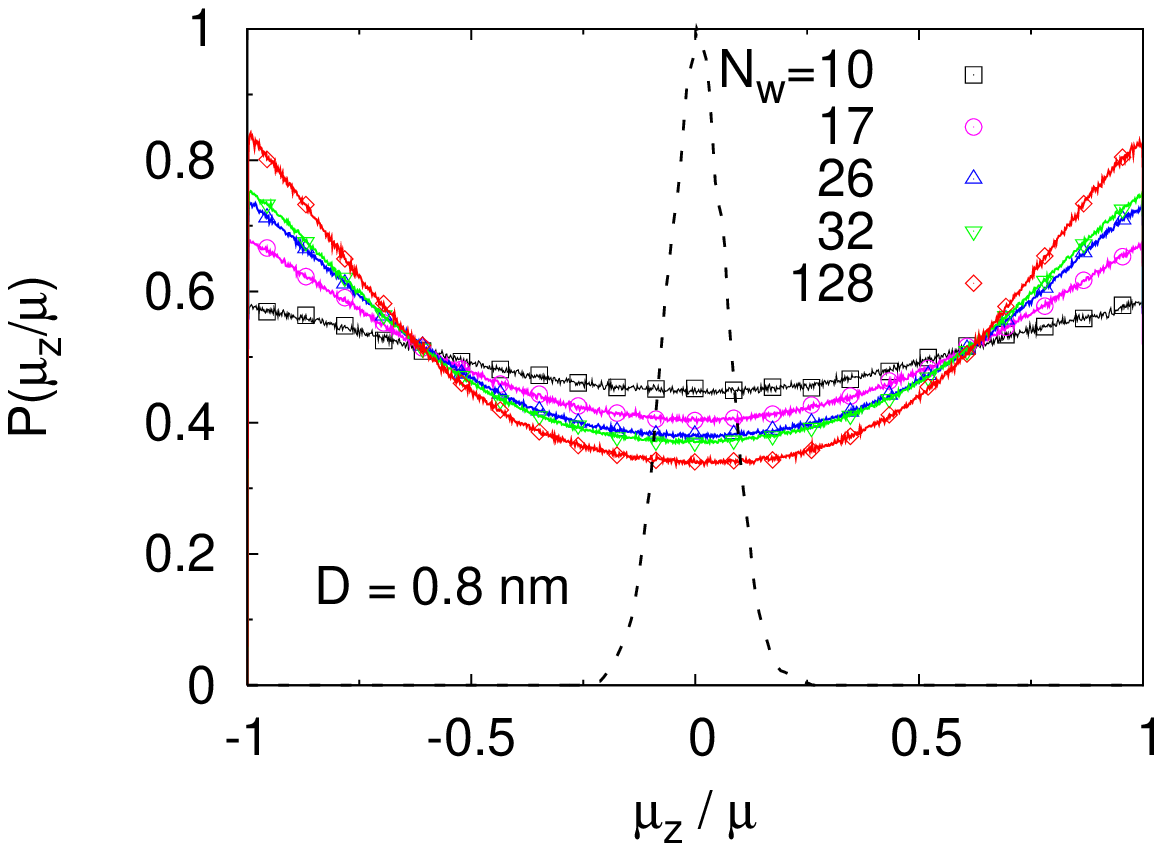}}
   \centerline{\includegraphics[width=5in]{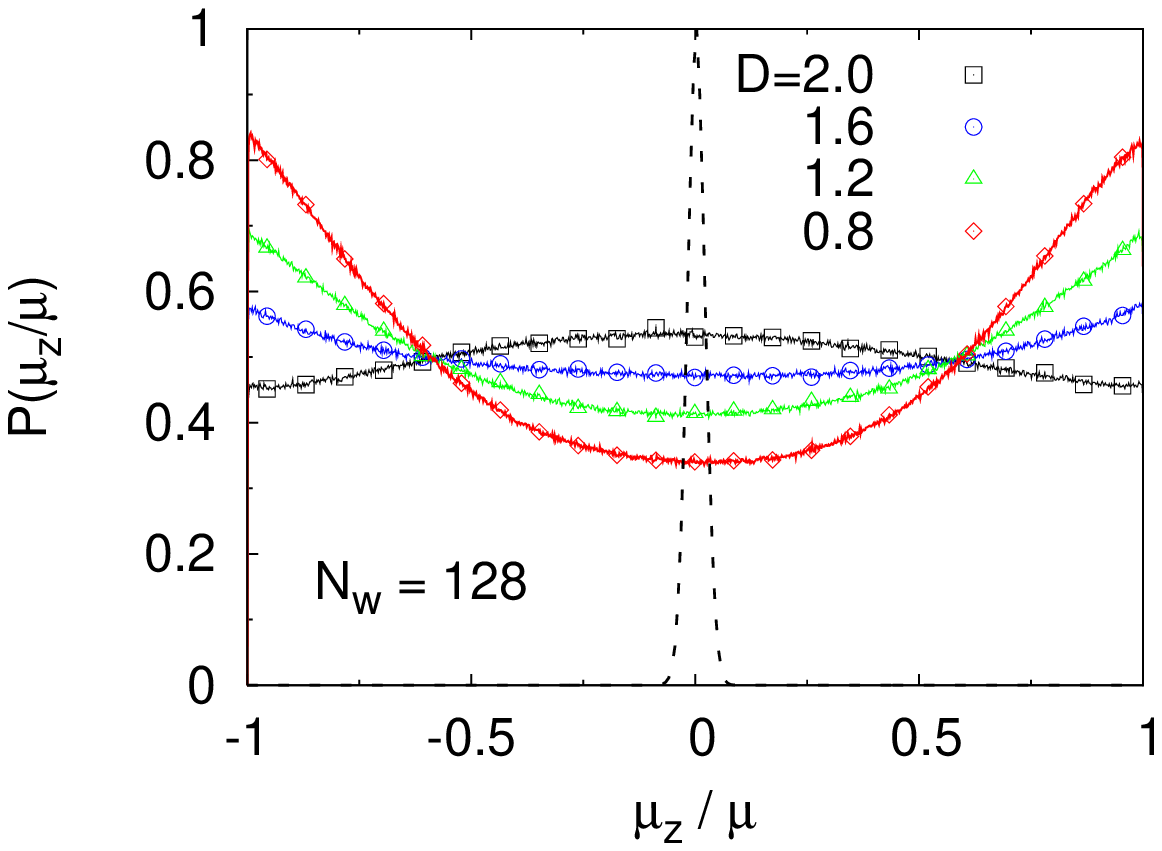}}
   \caption{}
   \label{fig:Mu}
 \end{figure}

Fig.~\ref{fig:Mu}: (color online) Probability distributions for $\mu_z/\mu$, where $\mu$ is the
  dipole moment of a single water molecule and $\mu_z$ is its component parallel
  to the pore axis. (a) with diameter $D=0.8$ nm and number of water molecules
  $N_w$ varying from 10 to 128 (b) with $N_w=128$ and $D$ varying from 0.8 to 2.0 nm.
  The dashed lines show the probability distributions for the total dipole
  moment of all water molecules in a nanopore with (a) $D=0.8$ nm and $N_w=128$
  and (b) $D=2.0$ nm and $N_w=128$.
  Water molecules are more aligned with the pore axis in long, narrow pores,
  but the net dipole moment is zero in all cases.

\begin{figure}[htbp]
   \centerline{\includegraphics[width=5in]{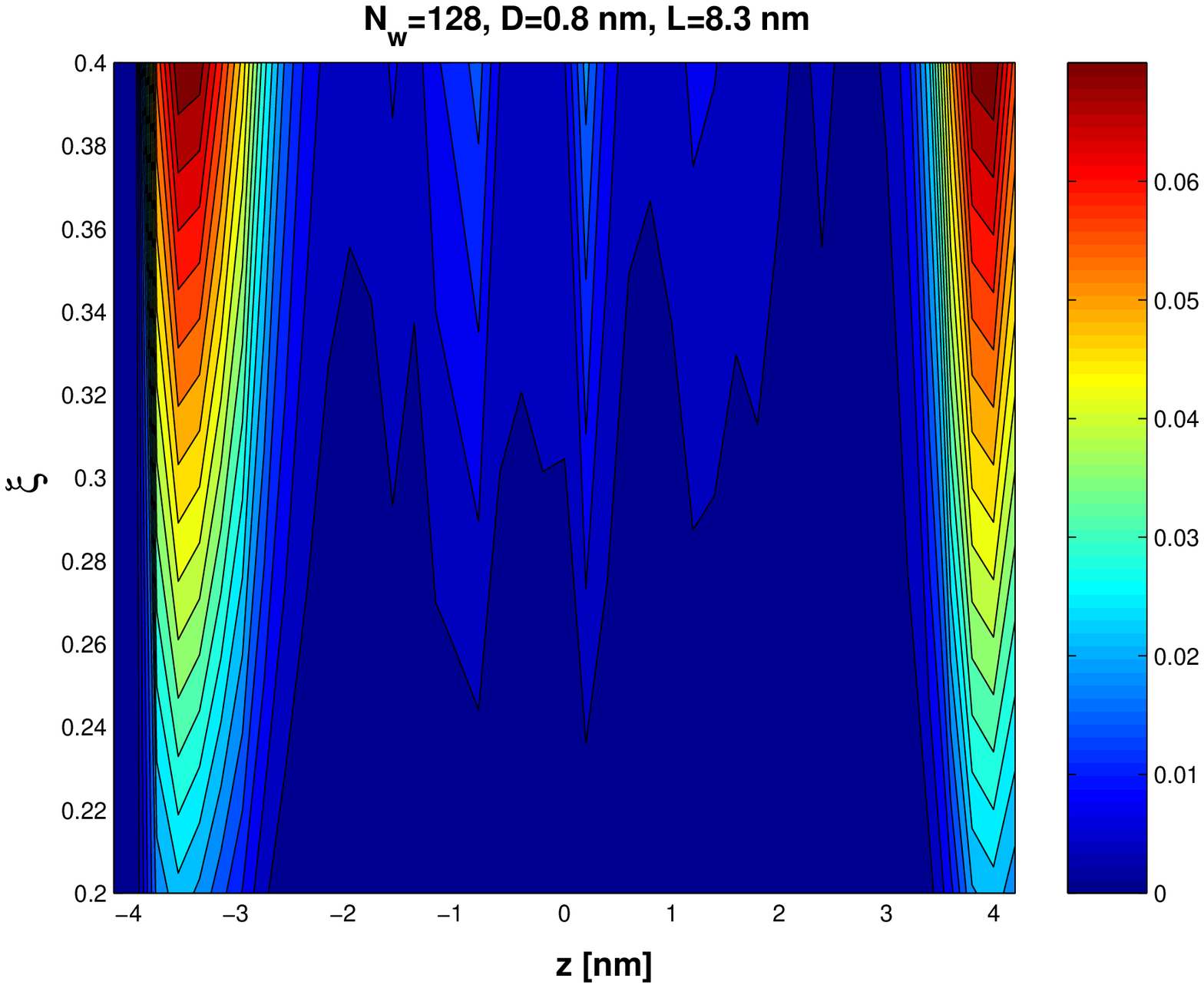}}
   \vspace{1cm}
   \centerline{\includegraphics[width=5in]{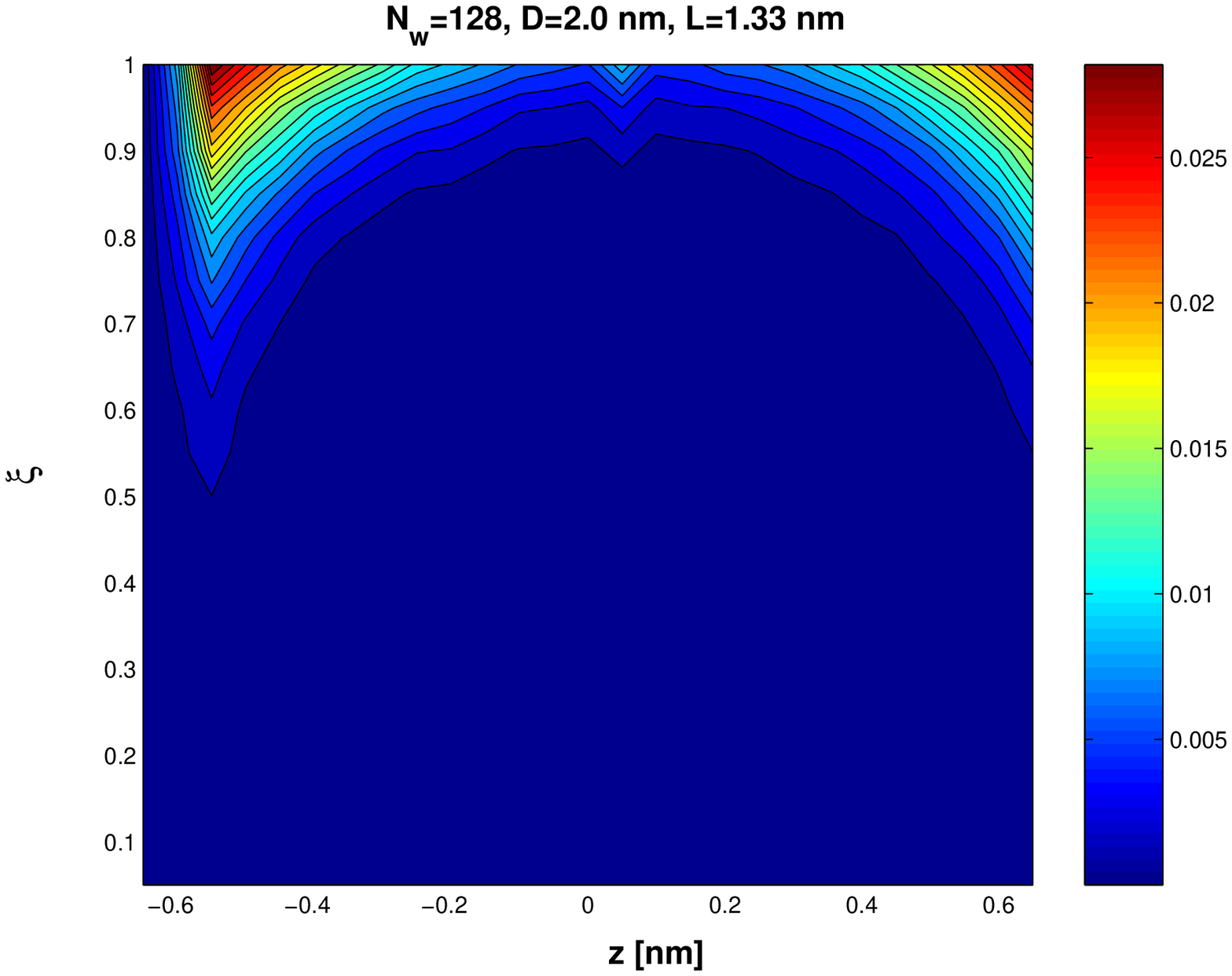}}
   \caption{}
   \label{fig:contour}
 \end{figure}

Fig.~\ref{fig:contour}: (color online) Contour plots of methane occupancies in nanopores with
  $N_w=128$ and (a) $D=0.8$~nm and (b) $D=2.0$ nm. The occupancy probabilities of
  the nonpolar solutes are greatest at the periphery of the pore.
  The scale for probability densities is shown on the right.

\begin{figure}[htbp]
   \centerline{\includegraphics[width=5in]{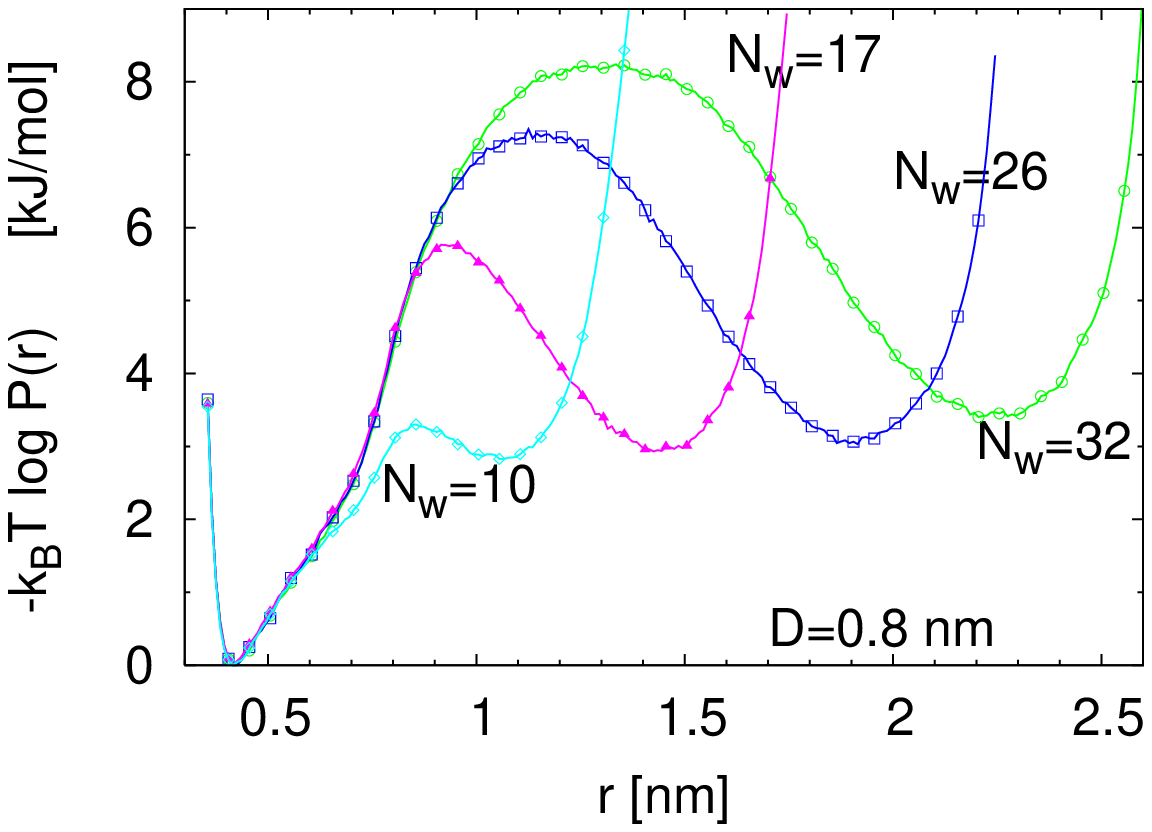}}
   \caption{}
   \label{fig:PMF_d0.8}
 \end{figure}

Fig.~\ref{fig:PMF_d0.8}: (color online) Free energies of interaction of two methane molecules
  in nanopores with diameter $D=0.8$ nm and number of water molecules $N_w$ varying from 10 to 32.
  All curves are translated vertically so that the reference state is at contact.
  The contact and one of the distant minima at $r \approx L-\sigma_{MM}$ are clearly resolved.
  The other distant minimum appears as a shoulder at $\sim0.7$ nm.

\begin{figure}[htbp]
   \centerline{\includegraphics[width=4in]{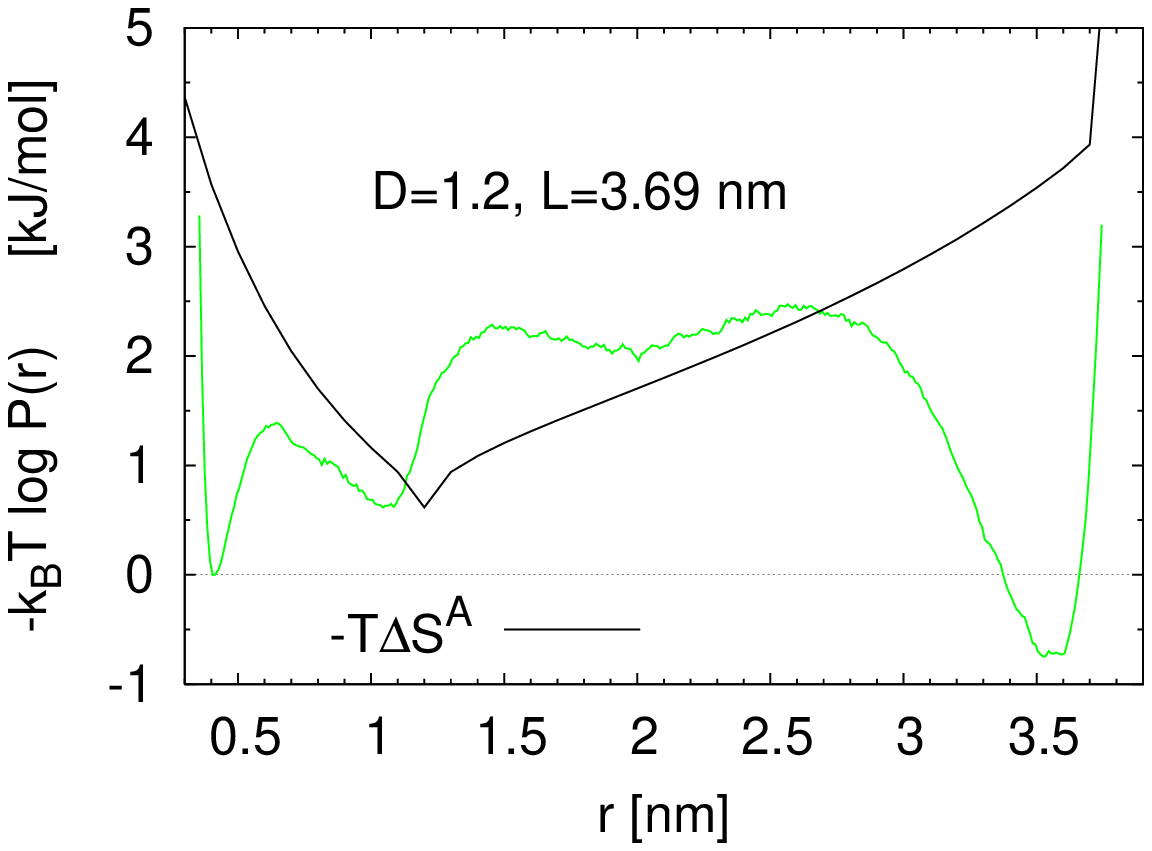}}
   \centerline{\includegraphics[width=4in]{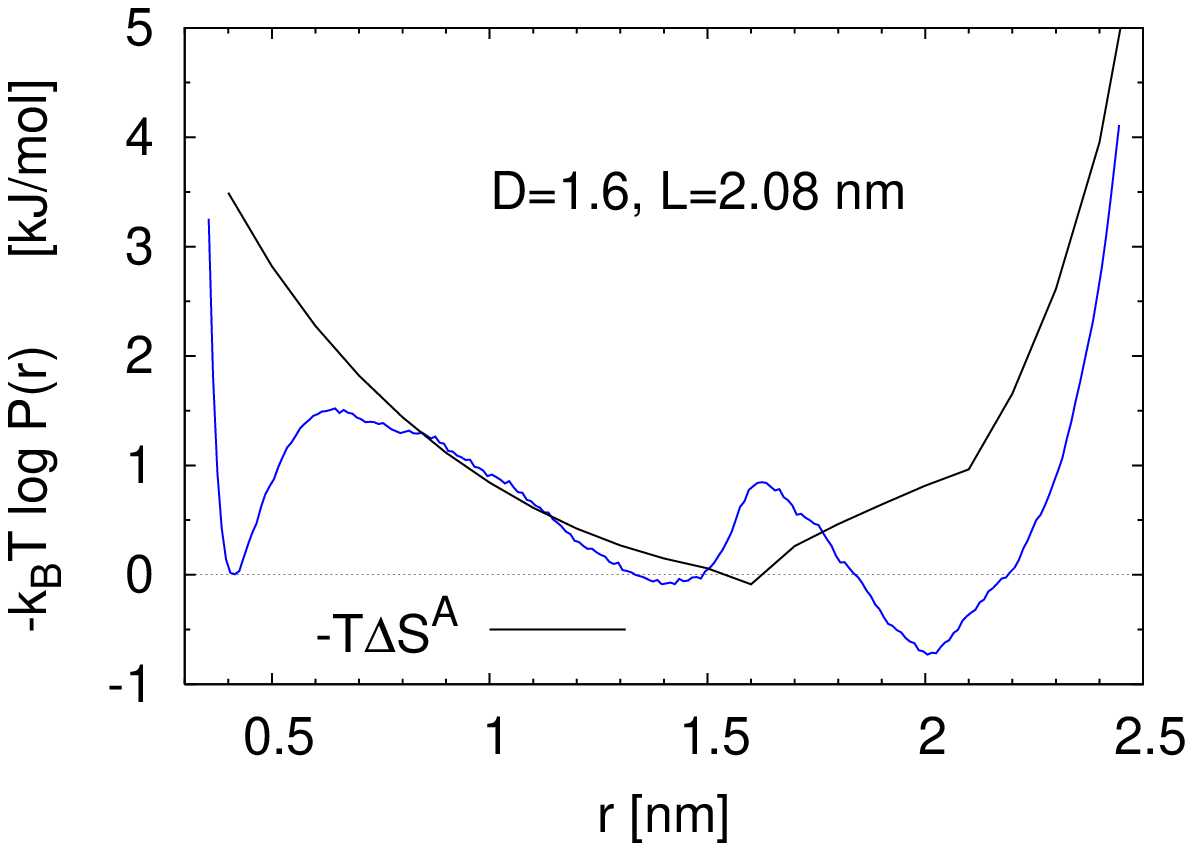}}
   \centerline{\includegraphics[width=4in]{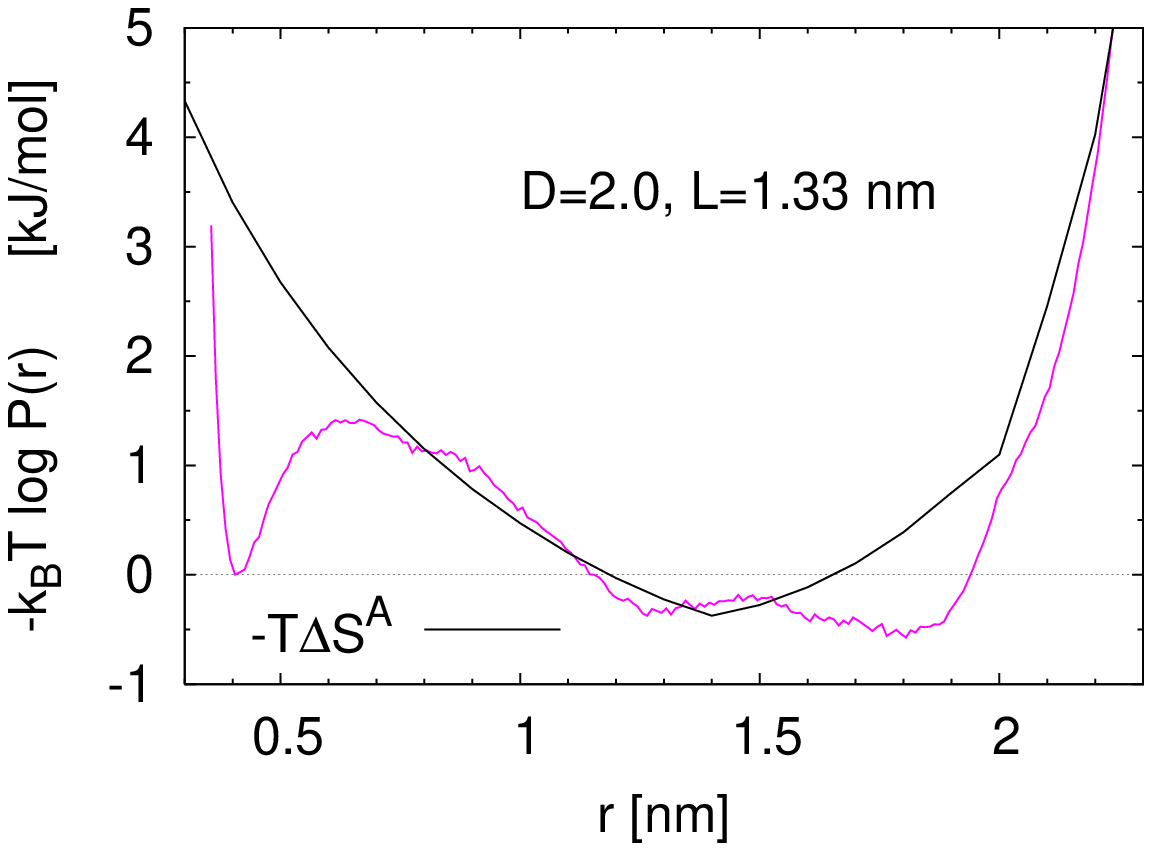}}
   \caption{}
   \label{fig:Trans_entropy}
 \end{figure}

Fig.~\ref{fig:Trans_entropy}:(color online) Interaction free energies of two methane molecules in
  nanopores with $N_w=128$ and pore diameter $D$ varying from 1.2 to 2.0 nm. The black lines
  show the contribution from the translational entropy of the hydrophobes, assuming that they
  are strictly confined to the surface.

\begin{figure}[htbp]
   \centerline{\includegraphics[width=5in]{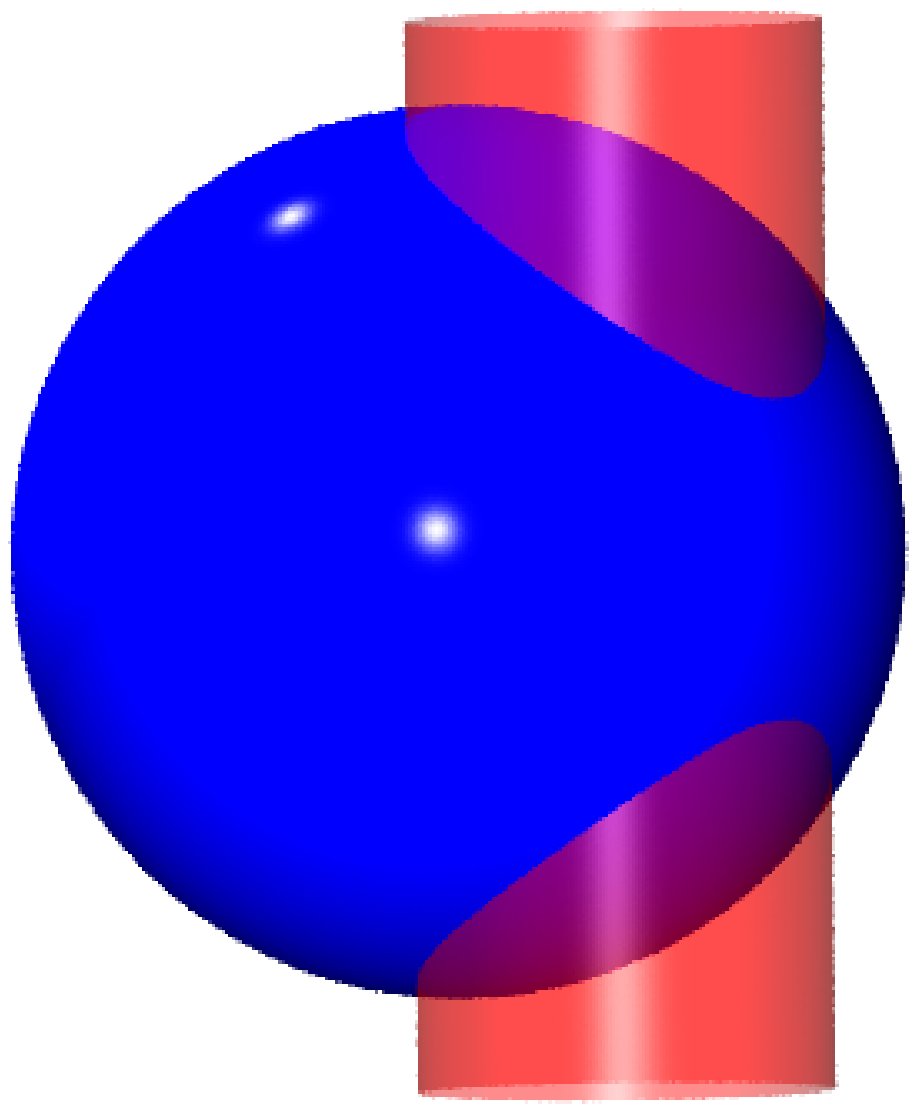}}
   \caption{}
   \label{fig:cyl_sph_inter}
 \end{figure}

Fig.~\ref{fig:cyl_sph_inter}: (color online) Illustration of the intersection of a cylinder
  and a sphere with its center on the body of the cylinder.

\begin{figure}[htbp]
   \centerline{\includegraphics[width=5in]{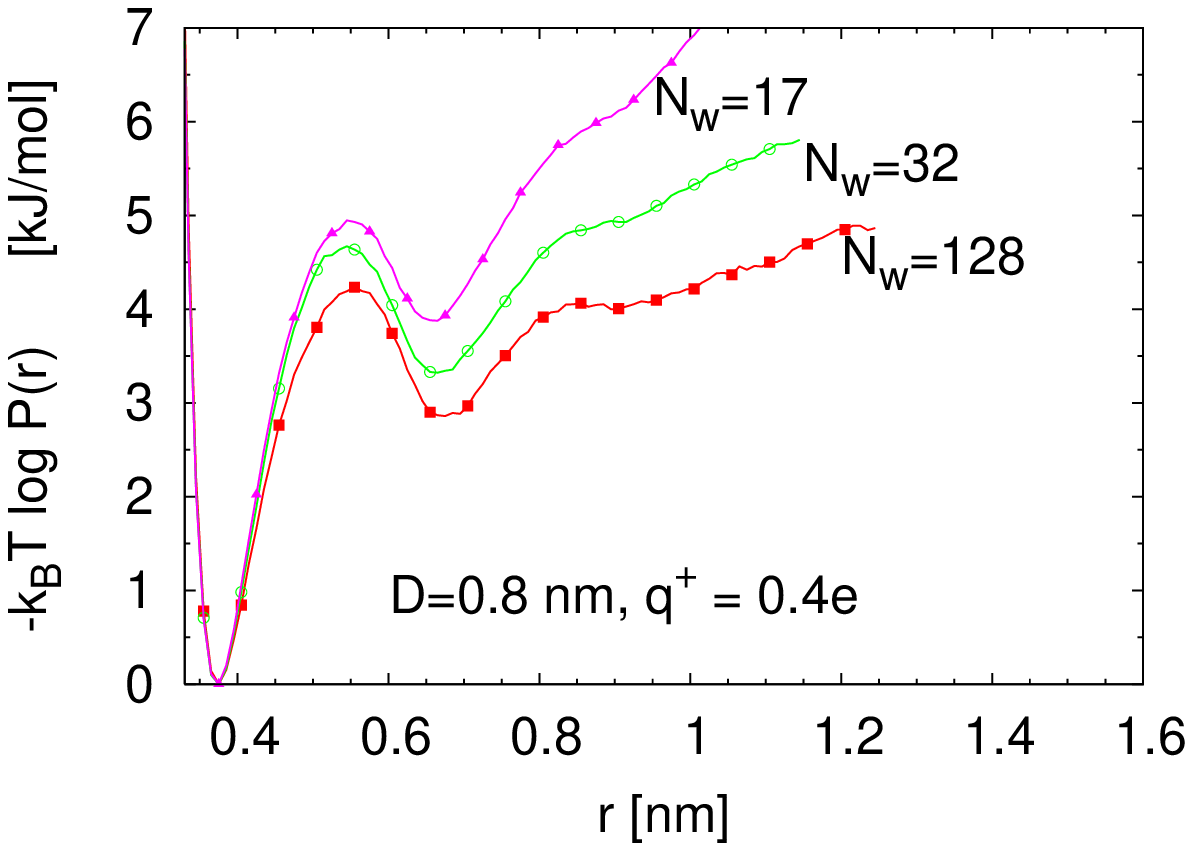}}
   \centerline{\includegraphics[width=5in]{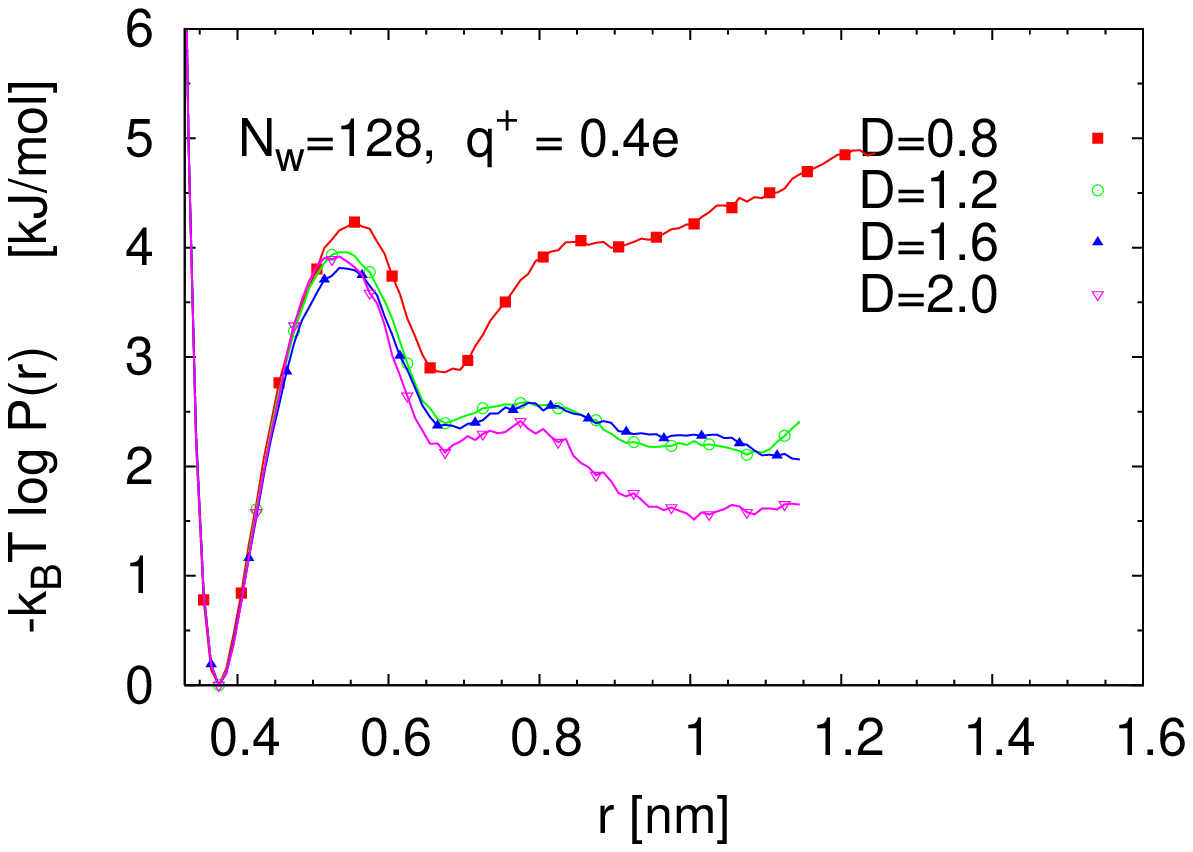}}
   \caption{}
   \label{fig:PMF_q0.4}
 \end{figure}

Fig.~\ref{fig:PMF_q0.4}:(color online) Free energies of interaction of model ions M$_{q+}$ and M$_{q-}$
  of charge magnitude $0.4e$ in nanopores (a) with diameter $D=0.8$ nm and number of water
  molecules $N_w$ varying from 17 to 128 (b) with $N_w=128$ and $D$ varying from 0.8 to 2.0
  nm. 

\begin{figure}[htbp]
   \centerline{\includegraphics[width=5in]{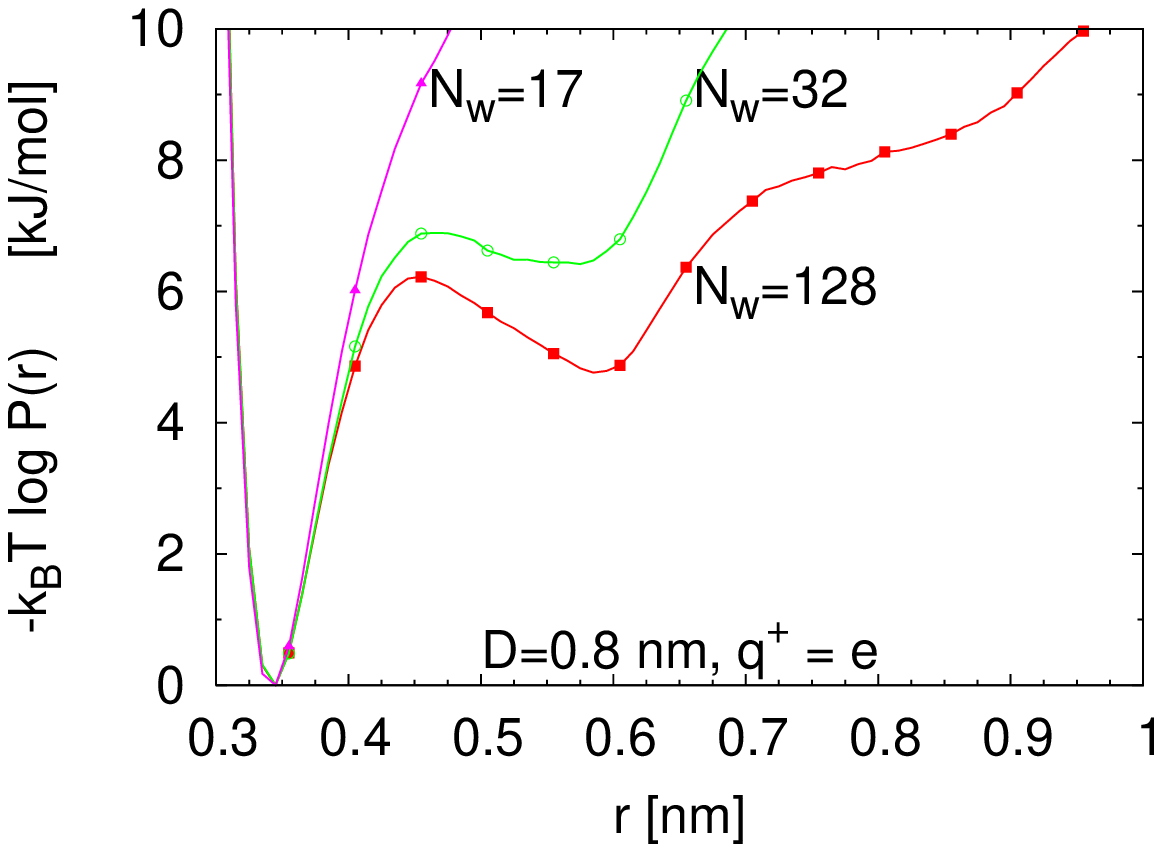}}
   \centerline{\includegraphics[width=5in]{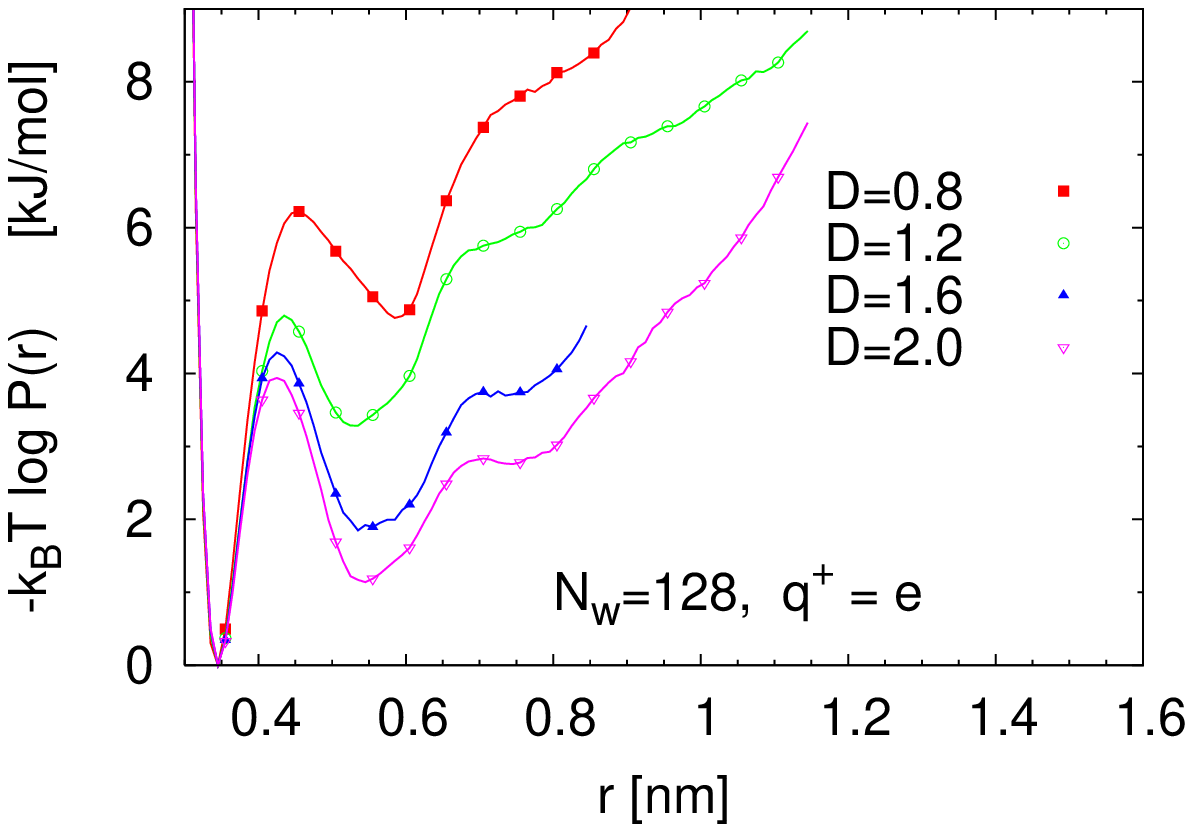}}
   \caption{}
   \label{fig:PMF_q1.0}
 \end{figure}

Fig.~\ref{fig:PMF_q1.0}:(color online) Free energies of interaction of model ions M$_{q+}$ and M$_{q-}$
  of charge magnitude $e$ in nanopores (a) with diameter $D=0.8$ nm and number of water
  molecules $N_w$ varying from 17 to 128 (b) with $N_w=128$ and $D$ varying from 0.8 to 2.0
  nm.

\end{document}